\documentclass[preprint,showpacs]{revtex4}
\usepackage{graphicx}
\usepackage{epsfig}
\input{epsf}

\begin{document}

\title{Quantitative Analysis of Photo-Thermal Stability of CdSe/CdS Core-Shell Nanocrystals}
\author{A.Singha}
\author{Anushree Roy}
\email{anushree@phy.iitkgp.ernet.in}
\affiliation{Department of Physics, Indian Institute of Technology,
Kharagpur 721 302, West Bengal, India}

\begin{abstract}
We report here investigations on the instability in luminescence
of bare (TOPO-stabilized) and CdS- capped CdSe particles under
infrared radiation. During photo-thermal annealing the formation
of oxide layers on the surfaces of the particles create defect
states. Consequently there is a reduction in particle size. These
two effects  control the light output from the samples. We make a
quantitative comparison of the stability of bare CdSe and
core-shell type CdSe-CdS particles under photo-annealing. Using
diffusion theory, we show that the volume of the oxide layer,
adhered to the crystallites, play a dominant role in controlling
the luminosity of the particles.


\end{abstract}

\maketitle

\section{Introduction}

Fluorescent nanoparticles have tremendous promise as indicators
and photon sources for a number applications such as biological
imaging, sensor technology, microarrays, and optical computing
\cite{Jaiswal:2003}. However, the practical applications of
nanocrystals demand efficiency and stability of the band edge
emission. It is well-known that the surface of nanocrystals is
made of atoms which are not fully coordinated. The unsaturated
dangling bonds are highly active and act like defects states
unless passivated. Due to high surface to volume ratio, the
contribution of  surface states is significant in controlling the
optical properties of nanoparticles. It is difficult to passivate
both cationic and anionic surface sites by long chain organic
materials \cite{Peng:1997}. For better surface passivation, the
core particles are usually capped with a higher band gap inorganic
compound. Nanoparticles with core-shell architecture have the
added benefit of providing a robust platform for incorporating
diverse functionalities into a single nanoparticle
\cite{Jackson:2001}. For CdSe nanocrystals, ZnS
\cite{Hines:1996,Kortan:1990} and CdS \cite{Tian:1996} are better
choices for capping because they (i) confine the photo-generated
charges in the core and (ii) have low lattice mismatch with core
material.

In the literature, we find several reports, where thermal and
photostability of such nanoparicles/nanorods have been discussed
\cite{Manna:2002, Peng:1997,Guo:2003}. For example, Manna \emph{et
al} \cite{Manna:2002}  have reported an increase in quantum
efficiency in CdS/ZnS nanorods due to photochemical annealing.
Photo-oxidation and photo-bleaching of single CdSe/ZnS dots have
been probed by room temperature time resolved spectroscopy
\cite{van Sark:2001}. In this article, the drop in luminosity of
the particles during thermal annealing has been explained by the
fall in number of emitted photons due to decrease in size of the
particles.The photochemical instability of capped CdSe
nanocrystals have also been explained by including photo-oxidation
of surface ligands followed by a precipitation process
\cite{Aldana:2001}. However, till date the detail quantitative
comparison of the time evolution of the light output from bare and
capped particles is unavailable.

In our work, we have addressed the above issues and discussed the
stability of TOPO-stabilized and inorganically passivated (CdS
capped) CdSe particles under Infrared (IR) radiation \emph{via}
photoluminescence (PL) measurements. The instability of the
particles during photo-thermal annealing in air have been
systematically and quantitatively studied till an asymptotically
constant behavior was observed with time.

\section{EXPERIMENTS}
The core-shell type CdSe-CdS and TOPO-stabilized CdSe particles
are prepared by following an established route
\cite{Pradhan:2003}. HDA and TOPO are melted at 60-80 $^{\circ}$C
and N$_{2}$ gas is bubbled into the melt for five minutes.
$7\times10^{-5}$ M SeUr is added to it at 100 $^{\circ}$C. In a
separate container $2\times10^{-4}$ M CdAc is dissolved in  HDA
followed by purging with nitrogen till a clear solution was
observed. The CdAc solution is rapidly added at 150 $^{\circ}$C to
the SeUr solution, which immediately results in an orange
coloration indicating the formation of CdSe particles (Sample A).
The temperature is slowly raised to 200 $^{\circ}$C and the
samples are collected at 70 $^{\circ}$C, followed by precipitation
by methanol. Decomposition of cadmium xanthate to CdS on the
surface of the CdSe formed the CdSe-CdS core-shell nanoparticles
(Sample B). For more detail of the sample preparation technique
one can refer to \cite{Pradhan:2003}.

To study the effect of photo-annealing, samples are kept under IR
radiation of temperature 60 $\pm$ 5 $^{\circ}$C. The PL spectra
are obtained using TRIAX 550 single monochromator with an open
electrode charge coupled detector(CCD). A 488 nm air cooled
Ar$^{+}$ ion laser with power 3 $\times$ 10$^3$ Watt/m$^2$ on the
sample is used as an excitation source .






\section{Photoluminescence and Thermal stability}

\begin{figure}
\centerline{\epsffile{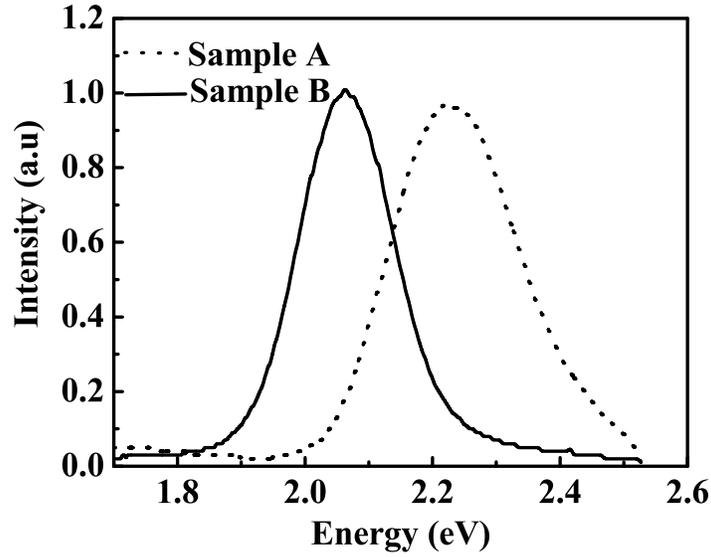}}
\caption{Photoluminescence
spectra of Sample A (dotted line) and Sample B (solid line)}
\end{figure}

The PL spectra of pristine Sample A (dotted) and Sample B (solid)
are shown in Fig. 1. Both the spectra are fitted with Gaussian
distribution function. From the best fit, the peak positions are
obtained as 557 nm (2.23 eV) for Sample A and 602 nm (2.06 eV) for
Sample B. The blue shift of the PL spectrum from the bulk
excitonic peak (712 nm) of CdSe is due to the confinement of
charge carriers in the nanostructures. From this energy shift and
using the well-known effective mass approximation (EMA) model
\cite{Efros:1982,Brus:1984}, the average particle size in Sample A
and Sample B are estimated to be 5.2 nm and 6.4 nm, respectively.

Subsequently, the samples are irradiated in air by IR lamp for 14
hours. The behavior of the luminescence spectra of the two samples
show a clear difference with evolution of time. For a detail
comparison of the stability of the particles in these samples
during photo-annealing we have fitted each PL spectrum, taken with
1.0 hr. interval (on average, more frequently in the beginning),
with the Gaussian function, keeping PL intensity, full width at
half maxima (FWHM) and peak position as fitting parameters. From
the obtained peak position the average size of the particles in
the samples has been estimated using the EMA model. The variation
of the particle size, PL spectral-width (FWHM) and peak intensity
with irradiation time for sample A and B are shown in Fig 2 and 3,
respectively. From Fig 2(a), 2(b) and 2(c) we see that PL spectrum
of Sample A exhibits a monotonic change with irradiation time : an
exponential degradation  of the average particle size and
luminescence intensity along with an exponential increase in the
spectral-width. However, the PL spectrum of Sample B exhibits a
non-monotonic behavior. For better understanding, we have divided
the total time of experiment for Sample B  into two regions [see
Fig 3(a) - 3(c)]. In region I (for about 1.5 hours) the
degradation of the particle size is a slow process. The
spectral-width remains almost constant and the luminescence
intensity increases with irradiation time. In region II (from 1.5
to 14 hours), the rate of degradation of the particle size and the
spectral-width is faster than what is observed in region I. The PL
intensity also decays exponentially in this region.

\begin{figure}
\centerline{\epsffile{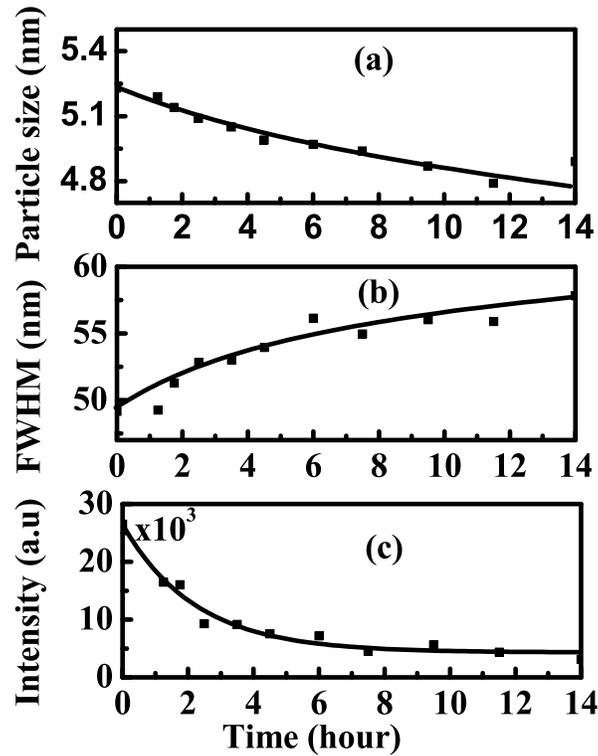}} \caption{Variation of (a)
particle size (b) FWHM and (c) luminescence intensity with time
for Sample A}.
\end{figure}

It is well-known that CdSe nanocrystals are highly photosensitive.
In our observation, the difference in the time evolution of the
emission spectra for TOPO-stabilized and CdS capped CdSe
nanocrystals is due to the difference in photo-oxidation process
in these two samples. TOPO is unstable under IR light and gets
detached from the particle surface. As a result, the unsaturated
bonds at the surface of bare CdSe particles without TOPO layer
directly oxidizes in light and air. Nanocrystals act as
photochemical catalysts for the oxidation of the free surface
ligands. It has been shown from the conductivity studies that the
oxygen is absorbed via a transfer of electron density from the
semiconductor to the oxygen \cite{katari:1994}. The main oxidation
products are CdSeO$_3$, SeO$_{2}$ and CdO . The oxygen proceeds to
break the chalcogen back bonds attached the bulk semiconductor,
from the surface to deep inside the particles \cite {katari:1994}.
Thus, with time the thickness of oxide layer increases and hence,
the diffusion of oxygen towards the centre of the particle (CdSe)
slows down. The aforementioned oxidation process and the change in
rate of diffusion of oxygen results in an exponential decrease in
the particle size with irradiation (Fig. 2a). The exponential drop
in PL intensity may be due to
 (i) the decrease in the number of photons emitted from the particles of smaller size and/or (ii)
 the formation of defect states at the interface of CdSe and oxide layer \cite{van Sark:2001}.
From Fig. 2 (b) we see that the above processes also influence the
size distribution of the particles, which increases the  width of
the PL spectrum.

\begin{figure}
\centerline{\epsffile{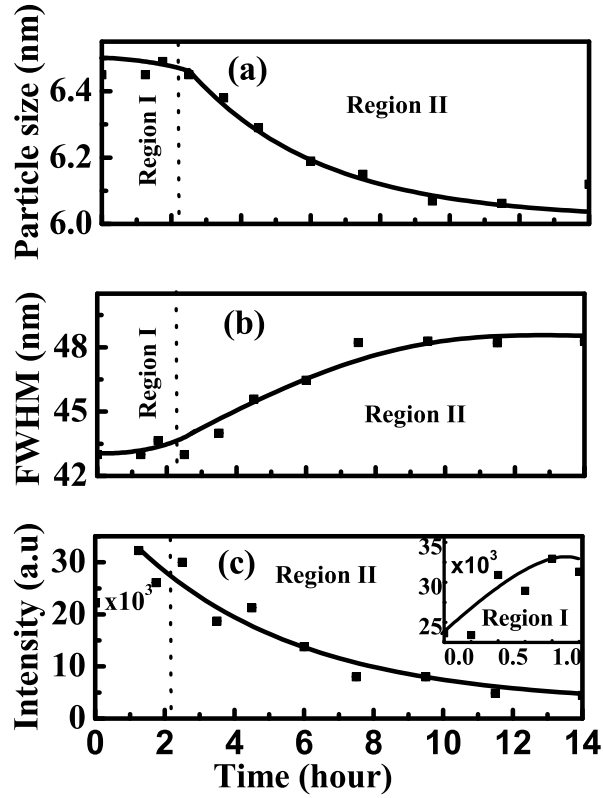}} \caption{Variation of (a)
particle size (b) FWHM and (c) and luminescence intensity with
time for Sample B. Inset of (c) shows the intensity variation for
first 1.5 hours}.
\end{figure}

For core-shell CdSe-CdS particles, oxygen has to diffuse through
the CdS shell. The slow variation in the particle size and
particle size distribution (reflected from the variation in
spectral-width) in region I of Fig. 3(a) and 3(b) indicate that
the oxygen diffuses slowly through thicker CdS layer. This is
supported by the fact that the oxidation of Se$^{2-}$ (as in case
of Sample A) is easier than that of S$^{2-}$ \cite{Cotton:1997}.
It has been shown that for core-shell particles the oxygen can
quench the defect states at the interface of core and shell
\cite{Henglein:1988, Chestnoy:1986}. A possible explanation of the
enhanced PL intensity in region I of Fig. 3(c) may be due to the
fact that the diffused oxygen in CdSe core through CdS layer is
\emph{just} sufficient to reduce the defect states at the
interface of core and shell without affecting the particle size
much. However, the prolonged exposure to light and heat in
presence of oxygen inevitably results in degradation of the
particles. The CdS is gradually photo-oxidized to Cd$^{2+}$ and
SO$_{4}^{2-}$\cite{van Sark:2001} and this causes a decrease in
shell thickness. Simultaneously, an oxide layer starts to grow
rapidly at core-shell interface. This gives rise to an exponential
decrease in the particle size and luminescence intensity in region
II of Fig 3(a) and 3(c) (similar to what was observed for Sample
A). We see that it is also reflected in the rapid increase of PL
spectral-width [region II of Fig.3(b)].
The photo-oxidation of the CdSe core through the CdS layer
indicates that the shells in the core-shell particles are not
closed epitaxial layers, rather they are layers with grain
boundaries. These grain boundaries are formed where the CdS
islands, which starts to grow at different places on the CdSe
nanocrystal surfaces, meet. At these boundaries the oxygen can
diffuse to the CdSe core inside the CdS shell.

\begin{table}[htbp]
\begin{tabular}{|c|c|c|} \hline
Sample   &  $\tau_{1}(hr)$  & $\tau_{2}(hr)$ \\ \hline
 A  &  4.2  &  2.0  \\ \hline
B  &  3.7 &  4.6  \\ \hline
\end{tabular}
\caption{The values of $\tau_1$ and $\tau_2$ for Sample A and
Sample B}
\end{table}

Now we proceed to analyze the results shown in  Fig. 2 and Fig. 3,
quantitatively. The exponential decay constants for sizes
($\tau_{1}$) of the particles and luminescence intensities
($\tau_{2}$) in Sample A and Sample B have been tabulated in Table
I. It is to be noted that though the values of $\tau$'s in Sample
B are nearly same; for Sample A the value of $\tau_1$ is two times
more than that of $\tau_2$.
Fick's law of diffusion for a diffusion of a species 'O' is given
by

\begin{equation}
\tau_{1}^{j}= - {D_{O}^{j}}\frac{d[O]}{dx},
\end{equation}
where $D_{O}^{j}$ is the diffusion coefficient of O and $d[O]/dx$
is the concentration gradient of O. The negative sign in Eq. 1
implies that diffusion takes place in a direction opposite to the
gradient. Since, we are are looking into the diffusion of oxygen
inside the CdSe particles of similar size, it is reasonable to
assume $d[O]/dx$ to be same for both Sample A and B
\cite{Aldana:2001}. We take into account the volume fraction,
$C_{col}^{j}$, of the oxide layer,which can be estimated from the
relation \cite{Signorini:2003}

\begin{equation}
C_{vol}^{j}(D)=\frac{\int_{0}^{\delta}\rho(z)^{j}(\frac{D}{2}-z)^2dz}{\int_{0}^{\delta}
(\frac{D}{2}-z)^2dz},
\end{equation}
where $\delta$ is the thickness of the oxide layer, which is
independent of the core diameter $D$. $\rho(z)^{j}$ is the
classical diffusion profile \cite{Signorini:2003}

\begin{equation}
\rho^{j}(z) = \exp[-(z/\lambda^{j})]^{2}
\end{equation}
Here, $\lambda^{j}$ is the length parameter which quantifies the
diffusion process and is proportional to the $D_{O}^{j}$ and hence
$\tau_{1}^{j}$ in Eq. 2 \cite{Signorini:2003}. In Eq. 1 - Eq. 3,
$j$ = A for Sample A and $j$ = B for Sample B. The average size of
the particles in the beginning ($d_i$) and also at the end ($d_f$)
of the experiment  and hence $\delta$ for Sample A and Sample B
can be easily estimated from Fig. 2(a) and 3(a). Using Eq. 2 the
ratio of $C_{vol}(D)$ in our two samples has been estimated by
taking
\begin{equation}
\frac{\lambda^{A}}{\lambda^{B}}=\frac{\tau_{1}^{A}}{\tau_{1}^{B}}=1.13
\end{equation}
All parameters, discussed above, are summarized Table II. We
calculate the ratio of the oxide volume ratio in Sample A and
Sample B to be 0.5, which is same as the ratio of the luminescence
decay rate in these samples (Table I).  It is now reasonable to
assume that the rather than size (which indirectly related to the
number of emitted photons), the volume of the oxide layer on the
surface of the particles plays a crucial role in determining the
stability of the luminescence output from the nanocrystals.

Schematic of the proposed mechanisms for the IR instability of the
samples is shown in Fig. 4. It is to be noted that nanocrystals
undergo Ostwald ripening only if the temperature of the system
rises beyond 100 $^{\circ}$, which is higher than what we use in
our experiment.

\begin{table}
\begin{tabular}{|c|c|c|c|c|c|c|c|} \hline
Sample   & d$_{i}$& d$_{f}$& $\delta$ & $\lambda$&
$\frac{C_{Vol}^{A}}{C_{Vol}^{B}}$ & Particle Vol. & Oxide Vol.\\
    & (nm) & (nm)& (nm) & ratio & & (nm$^3$) &ratio\\\hline
 A  &  5.23 &  4.78 & 0.45 & 1.13 & 1 & 57.1 & 0.5\\
B  &  6.45 &  6.12 & 0.33 &      &   & 120 &  \\ \hline
\end{tabular}
\caption{Parameters used from Fig. 2 and Fig. 3 to estimate the
ratio of volume fraction and oxide-volume in Sample A and Sample
B}
\end{table}

\section{CONCLUSION}
\begin{figure}
\centerline{\epsffile{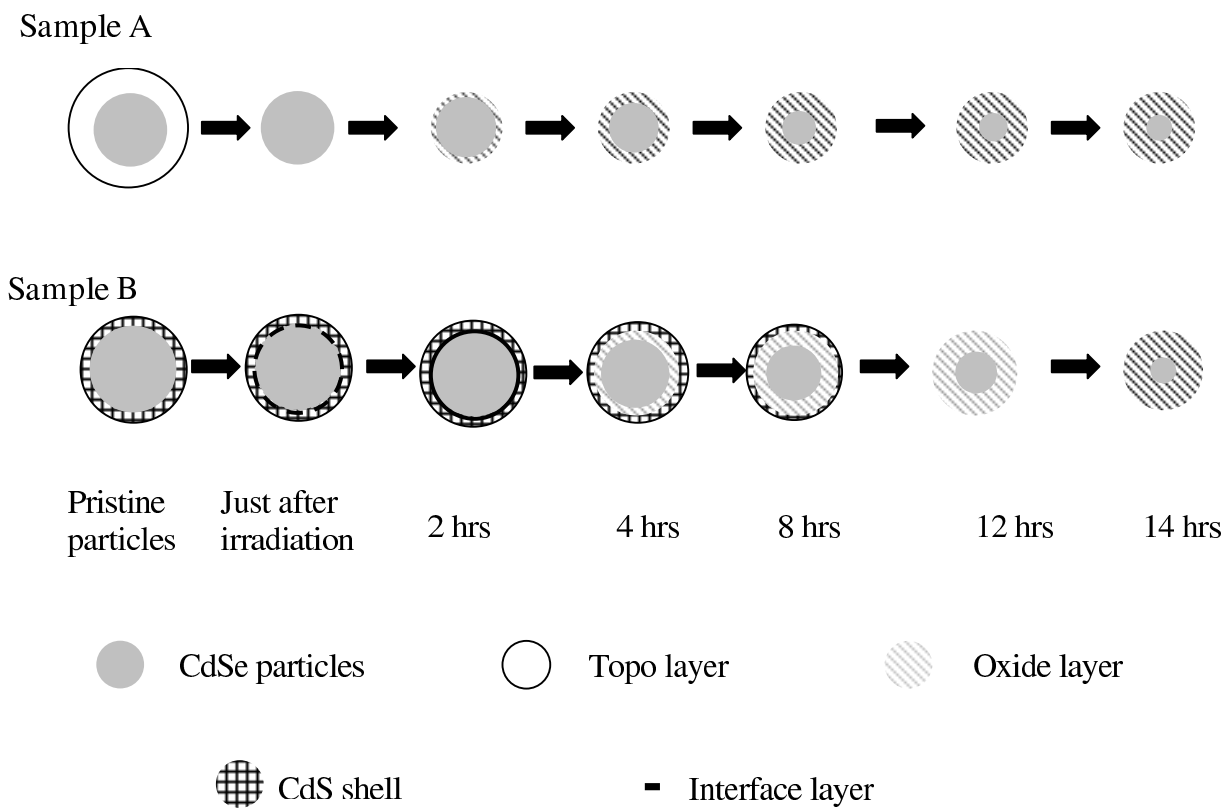}} \caption{Schematic of the
proposed mechanisms for the IR instability of the samples}
\end{figure}

In this article, we have quantitatively analyzed the stability of
bare and CdS capped core-shell particles under IR radiation.
Radiation causes photo-annealing of the samples.  The decay rate
of the light output from CdS capped particles is found to be two
times slower than that of bare particles. Using a simple diffusion
model we conclude that rather than size of the particles, the
volume of the oxide layer surrounding the particles play an
important role in controlling their luminosity. It is worthwhile
to check the above model for the particles with different shell
thickness.

\begin{acknowledgements}
Authors thank Department of Science and Technology, India for
financial support. Authors also thank N. Pradhan for providing the
samples.
\end{acknowledgements}

\end{document}